\documentclass[aps,pra,showkeys,twocolumn,superscriptaddress]{revtex4-2}
\usepackage{array}
\usepackage{booktabs}
\usepackage{tabu}
\usepackage{dcolumn}
\usepackage{amsmath}
\usepackage{amsfonts}
\usepackage{float}
\usepackage{amssymb}
\usepackage{graphicx,color}
\usepackage[colorlinks={true}]{hyperref}
\hypersetup{colorlinks=true,linkcolor=red,citecolor=blue,urlcolor=blue}
\usepackage{graphicx}
\usepackage{subfigure}
\usepackage{graphicx}% Include figure files
\usepackage{dcolumn}% Align table columns on decimal point
\usepackage{bm}% bold math
\usepackage{pstricks}
\usepackage{braket}
\usepackage{orcidlink}

\def\be{\begin{equation}}
  \def\ee{\end{equation}}
\def\bea{\begin{eqnarray}}
\def\eea{\end{eqnarray}}
\def\f{\frac}
\def\n{\nonumber}
\def\l{\label}
\def\p{\phi}
\def\o{\over}
\def\R{\rho}
\def\pa{\partial}
\def\om{\omega}
\def\na{\nabla}
\def\P{\Phi}
\begin{document}

\title{Tripartite measurement uncertainty in Schwarzschild space-time}

\author{Hazhir Dolatkhah\orcidlink{0000-0002-2411-8690}}
\email{h.dolatkhah@gmail.com}
\affiliation{RCQI, Institute of Physics, Slovak Academy of Sciences, \\D\'{u}bravsk\'{a} cesta 9, 84511 Bratislava, Slovakia}

\author{Artur Czerwinski\orcidlink{0000-0003-0625-8339}}
\affiliation{Institute of Physics, Faculty of Physics, Astronomy and Informatics, Nicolaus Copernicus University in Torun, ul. Grudziadzka 5, 87–100 Torun, Poland}

\author{Asad Ali\orcidlink{0000-0001-9243-417X}}
\affiliation{Qatar Centre for Quantum Computing, College of Science and Engineering, Hamad Bin Khalifa University, Doha, Qatar}

\author{Saif Al-Kuwari\orcidlink{0000-0002-4402-7710}}
\affiliation{Qatar Centre for Quantum Computing, College of Science and Engineering, Hamad Bin Khalifa University, Doha, Qatar}

\author{Saeed Haddadi\orcidlink{0000-0002-1596-0763}} \email{haddadi@semnan.ac.ir}
\affiliation{Faculty of Physics, Semnan University, P.O. Box 35195-363, Semnan, Iran}
\affiliation{Saeed's Quantum
	Information Group, P.O. Box 19395-0560, Tehran, Iran}

\date{\today}% It is always \today, today,

%%%%%%%%%%%%%%%%%%%%%
%%%%%%%%%%%%%%%%%%%%
%%%%%%%%%%%%%%%%%%%%%%
%%%%%%%%%%%%%%%%%%%%%%%%

\def\be{\begin{equation}}
  \def\ee{\end{equation}}
\def\bea{\begin{eqnarray}}
\def\eea{\end{eqnarray}}
\def\f{\frac}
\def\n{\nonumber}
\def\l{\label}
\def\p{\phi}
\def\o{\over}
\def\R{\rho}
\def\pa{\partial}
\def\om{\omega}
\def\na{\nabla}
\def\P{\Phi}
%\nofiles

%=============================================================%
%=============================================================%
%============== Abstract =======================================%
%=============================================================%
%=============================================================%
\begin{abstract}
\textbf{Abstract.} The effect of Hawking radiation on tripartite measurement uncertainty in a Schwarzschild black hole background is analyzed in this study. Two scenarios are examined. In the first, quantum memory particles approach a Schwarzschild black hole and are positioned near the event horizon, while the particle being measured remains in the asymptotically flat region. In the second scenario, the measured particle moves toward the black hole, and the quantum memories stay in the asymptotically flat region. This study considers two initial quantum states, namely GHZ and W states. Our findings reveal that in both cases, measurement uncertainty increases steadily with rising Hawking temperature. When comparing the GHZ and W states, the GHZ state initially exhibits lower measurement uncertainty at low Hawking temperatures than the W state, indicating greater resilience to Hawking radiation. Additionally, when the quantum memories remain in the asymptotically flat region while the measured particle falls toward the black hole, the uncertainties for GHZ and W states do not align at high temperatures. The GHZ state consistently demonstrates lower measurement uncertainty, showcasing its superior robustness against Hawking radiation.
\end{abstract}
%\pacs{04.50}

\maketitle

%%%%%%%%%%%%%%%%%%%%%%%%%%%%%%%%%%%%%%%%%%%%%%%%%%%%%%%%%%%%%%%%%%%%%%%%%%%%
%%%%%%%%%%%%%%%%%%%%%%%%%%%%%%%%%%%%%%%%%%%%%%%%%%%%%%%%%%%%%%%%%%%%%%%%%%%%
%%%%%%%%%%%%%%%%%%%%%%%%%%%%%%%%%%%%%%%%%%%%%%%%%%%%%%%%%%%%%%%%%%%%%%%%%%%%
%%%%%%%%%%%%%%%%%%%%%%%%%%%%%%%%%%%%%%%%%%%%%%%%%%%%%%%%%%%%%%%%%%%%%%%%%%%%
%============  Sec.I (Introduction)  =======================================
%%%%%%%%%%%%%%%%%%%%%%%%%%%%%%%%%%%%%%%%%%%%%%%%%%%%%%%%%%%%%%%%%%%%%%%%%%%%
%%%%%%%%%%%%%%%%%%%%%%%%%%%%%%%%%%%%%%%%%%%%%%%%%%%%%%%%%%%%%%%%%%%%%%%%%%%%
%%%%%%%%%%%%%%%%%%%%%%%%%%%%%%%%%%%%%%%%%%%%%%%%%%%%%%%%%%%%%%%%%%%%%%%%%%%%
%%%%%%%%%%%%%%%%%%%%%%%%%%%%%%%%%%%%%%%%%%%%%%%%%%%%%%%%%%%%%%%%%%%%%%%%%%%%
\section{Introduction}	%) A SECTION HEADING
The uncertainty principle can be represented using Shannon entropy \cite{Deutsch}. Specifically, it has been demonstrated that for two incompatible observables $X$ and $Z$, the following entropic uncertainty relation (EUR) applies \cite{Uffink}
\begin{equation}\label{Maassen and Uffink}
H(X)+H(Z)\geqslant -\log_2 c,
\end{equation}
where $H(P) = -\sum_{k} p_k \log_2 p_k$ represents the Shannon entropy of the measured observable $P \in \lbrace X, Z \rbrace$, in which $p_k$ is the probability of obtaining measurement outcome of $k$. The term $c$ measures the complementarity between the observables and is defined as $c = \max_{\lbrace \mathbb{X},\mathbb{Z}\rbrace } \vert\langle x_{i} \vert z_{j}\rangle \vert ^{2}$, with $\mathbb{X}=\lbrace \vert x_{i}\rangle \rbrace$ and $\mathbb{Z}=\lbrace \vert z_{j}\rangle \rbrace$ being the eigenbases of the observables $X$ and $Z$, respectively.

In the last decade, many efforts have been made to generalize and modify this relation \cite{Berta,Renes,Coles1,Bialynicki,Wehner,Pati,Ballester,Vi,Wu,Rudnicki,Pramanik,Maccone,Coles,Adabi,Dolat2,Dolatkhah,Zozor,R,Kamil,Rudnicki1,Pramanik1,Ming,Dolat,hadsci,wangpra2021,wangpra2022,dolatkhahepjp,Renpra2023,wangplb2024}.
Uncertainty relations can be understood through a tripartite scenario illustrated by an uncertainty game involving three participants: Alice, Bob, and Charlie. At the start of the game, Alice, Bob, and Charlie share a quantum state $\rho_{ABC}$. In the next step, Alice performs one of two possible measurements, $X$ or $Z$, and then informs Bob and Charlie, who hold the quantum memories $B$ and $C$ respectively, about her measurement choice. If $X$ is measured, Bob's task is to predict the outcome of this measurement. If $Z$ is measured, then Charlie's task is to predict the outcome of Alice's measurement. It has been demonstrated that the tripartite quantum memory entropic uncertainty relation (QM-EUR) can be formulated as \cite{Renes, Berta}
\begin{equation}\label{tpu}
U\equiv S(X \vert B)+S(Z \vert C)\geqslant -\log_2 c,
\end{equation}
where $c$ is identical to the term defined in Eq. \eqref{Maassen and Uffink}. Besides, $S(X \vert B) $ and $S(Z \vert C) $ represent the conditional von Neumann entropies of the states
\begin{equation}
\rho_{XB}= \sum_{i}(\vert x_{i}\rangle_{A}\langle x_{i}\vert\otimes \mathbf{I}_B ) \rho_{AB}(\vert x_{i}\rangle_{A}\langle x_{i}\vert\otimes \mathbf{I}_{B} ),
\end{equation}
 and
\begin{equation}
\rho_{ZC}= \sum_{j}(\vert z_{j}\rangle_{A}\langle z_{j}\vert\otimes \mathbf{I}_{C} ) \rho_{AC}(\vert z_{j}\rangle_{A}\langle z_{j}\vert\otimes \mathbf{I}_{C} ),
\end{equation}
 respectively, where $\mathbf{I}_{B(C)}$ is the identity operator in Bob (Charlie)’s Hilbert space.

 Physically speaking, $S(X \vert B) $ quantifies Bob’s uncertainty about the outcome of measurement $X$  given that Bob has access to the quantum memory $B$ and, likewise, $S(Z \vert C) $ quantifies Charlie's uncertainty about the outcome of measurement $Z$ given that Charlie has access to the quantum memory $C$.

The study of the influence of the relativistic effect on quantum correlations in curved space-time has been the main topic of many researches \cite{black1,black2,black3,black4,black5,black6,black7,black8,wangepjc1,wangepjc2,wangprb2024,black9,black10}. Especially, the influence of Hawking radiation on quantum entanglement is studied in the background of a Schwarzschild black hole \cite{black9,black10}. It has been shown that in curved space-time, the presence of Hawking radiation can reduce quantum correlations \cite{black9,black10,new1,new2,new3}. Also, it is well known that quantum correlations between the quantum memory and the measured particle play an important role in QM-EURs. Note that quantum entanglement between the memory particle and the measured particle can decrease uncertainty. Therefore, it is important to examine how the Hawking effect impacts the tripartite QM-EUR in the context of black holes. While the Hawking effect on QM-EUR has been thoroughly explored in bipartite systems \cite{black11,black12,black13,black14}, to the best of our knowledge, the tripartite QM-EUR has been investigated in only a few studies such as Ref. \cite{black15}.

Driven by these observations, we explore the impact of Hawking radiation on the tripartite QM-EUR within Schwarzschild space-time. In our study, we assume that Alice, Bob, and Charlie initially share a generally tripartite entangled state in flat Minkowski space-time. In the next step, we consider two different scenarios. In the first scenario, Bob and Charlie, who hold the quantum memories $B$ and $C$, freely fall towards a Schwarzschild black hole and position themselves near the event horizon, while Alice stays in the asymptotically flat region. In the second scenario, Alice freely falls toward a Schwarzschild black hole and then hovers near the event horizon, while Bob and Charlie remain in the asymptotically flat region. We examine the behavior of the tripartite QM-EUR in these scenarios for two distinct classes of tripartite entangled states, i.e. the Greenberger–Horne–Zeilinger (GHZ) state and the W state.

The structure of this paper is as follows. In Sec. \ref{Sec2}, we provide a brief overview of the vacuum structure and Hawking radiation for Dirac fields in Schwarzschild space-time. In Sec. \ref{Sec3}, we analyze the impact of the Hawking effect on the tripartite QM-EUR in Schwarzschild space-time. The final section is dedicated to the conclusion and discussion.

%%%%%%%%%%%%%%%%%%%%%%%%%%
%%%%%%%%%%%%%%%%%%%%%%%%%%%
%%%%%%%%%%%%%%%%%%%%%%%%%%
%%%%%%%%%%%%%%%%%%%%%%%%%%
\section{Dirac fields in Schwarzschild space-time}\label{Sec2}
The metric for Schwarzschild space-time can be written as follows:
\begin{align}
d s^2= &-\left(1-\frac{2 M}{r}\right) d t^2+\left(1-\frac{2 M}{r}\right)^{-1} d r^2\nonumber\\
&+r^2\left(d \theta^2+\sin ^2 \theta d \varphi^2\right),
\end{align}
where $M$ represents the mass of the black hole. Throughout this paper, we assume that $G= c= \hbar=\kappa_B=1$.

The Dirac equation \cite{Dirac} can be formulated as $\left[\gamma^a e_a^\mu\left(\partial_\mu+\Gamma_\mu\right)\right] \psi=0$, where $\gamma^a$ denotes the Dirac matrix, $e_a^\mu$ is the inverse tetrad, and $\Gamma_\mu$ represents the spin connection coefficient. For Schwarzschild space-time, the Dirac equation specifically takes the form:
{\small \begin{align}
& -\frac{\gamma_0}{\sqrt{1-\frac{2 M}{r}}} \frac{\partial \psi}{\partial t}+\gamma_1 \sqrt{1-\frac{2 M}{r}}\left[\frac{\partial}{\partial r}+\frac{1}{r}+\frac{M}{2 r(r-2 M)}\right] \psi \nonumber\\ &+\frac{\gamma_2}{r}\left(\frac{\partial}{\partial \theta}+\frac{\cot \theta}{2}\right) \psi+\frac{\gamma_3}{r \sin \theta} \frac{\partial \psi}{\partial \phi}=0.
\end{align}}
By solving the above equation, one can obtain the positive (fermions) frequency outgoing solutions for the outside and inside regions of the event horizon \cite{Dirac1}
\begin{equation}
\Psi^{I}_{k}=\xi e^{-i\omega u} \quad \textmd{and} \quad \Psi^{II}_{k}=\xi e^{+i\omega u},
\end{equation}
 respectively, where  $k$  represents the wave vector, $\xi$  is a four-component Dirac spinor, and $\omega$ is a monochromatic frequency of the Dirac field. The retarded time $u$  is expressed as $u= t-r_{\ast}$ in which $r_{\ast}=r+2M \ln \frac{r-2M}{2M}$  is the tortoise coordinate.\\

Making an analytic extension for the above equation through Damour and Ruffini’s suggestion \cite{Dirac2}, one can provide a complete basis for the positive energy modes. Subsequently, one can obtain the Bogoliubov transformations \cite{Dirac3} between the creation and annihilation operators in the Schwarzschild and Kruskal coordinates by quantizing the Dirac fields in the Schwarzschild and Kruskal modes, respectively. After suitably normalizing the state vector,  the expressions of the Kruskal vacuum and excited states can be expressed as
 \begin{equation}\label{bl1}
 \vert 0\rangle_{k}=\alpha \vert 0\rangle^{+}_{I} \vert 0\rangle^{-}_{II}+\beta \vert 1\rangle^{+}_{I} \vert 1\rangle^{-}_{II}
 \end{equation}
 and
 \begin{equation}\label{bl2}
 \vert 1\rangle_{k}= \vert 1\rangle^{+}_{I} \vert 0\rangle^{-}_{II},
 \end{equation}
where $ \alpha=\frac{1}{\sqrt{e^{-\omega/T}+1}}$, $ \beta=\frac{1}{\sqrt{e^{\omega/T}+1}}$, and  $ T=\frac{1}{8\pi M}$ is the Hawking temperature \cite{Dirac4}. Here, $\vert n\rangle^{+}_{I}$  and $\vert n\rangle^{-}_{II}$ correspond to the orthonormal bases for the fermion in the outside region and the antifermion in the inside region of the event horizon. In the following,  both  $\vert n\rangle^{+}_{I}$  and $\vert n\rangle^{-}_{II}$ will be denoted as  $\vert n\rangle_{I}$  and $\vert n\rangle_{II},$ in order to simplify the formulation.

\begin{figure}[t]
	\begin{center}     \includegraphics[width=0.43\textwidth]{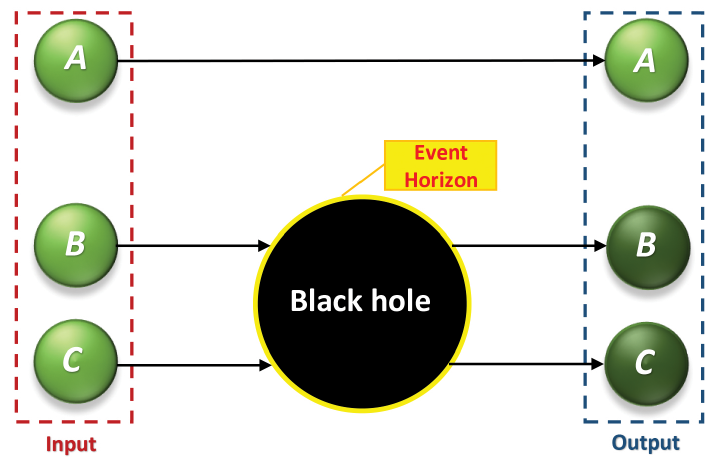}	
	\end{center}
	\caption{A schematic diagram of our model shows Alice's particle, $A$, located in a flat region, while Bob's particle, $B$, and Charlie's particle, $C$, are positioned near the event horizon of a Schwarzschild black hole. The dashed lines represent the entanglement between the particles. The input states are given in \eqref{GHZ} and \eqref{W}, and the corresponding output states are presented in \eqref{GHZ2} and \eqref{W2}.}
	\label{Schematic1}
\end{figure}

\begin{figure}[t]
	\begin{center}     \includegraphics[width=0.43\textwidth]{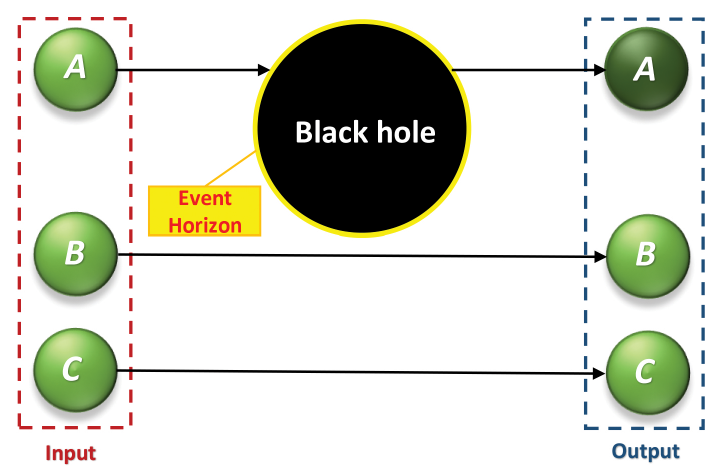}	
	\end{center}
	\caption{This schematic diagram illustrates the model with Alice's particle, $A$, positioned near the event horizon of a Schwarzschild black hole, while Bob's particle, $B$, and Charlie's particle, $C$, are located in a flat region. The input states are shown in \eqref{GHZ} and \eqref{W}, with the corresponding output states provided in \eqref{GHZ3} and \eqref{W3}.}
	\label{Schematic2}
\end{figure}

%%%%%%%%%%%%%%%%%%%%%%%%%%
%%%%%%%%%%%%%%%%%%%%%%%%%%%%%%%
%%%%%%%%%%%%%%%%%%%%%%%%%%%%%%%
\section{Results}\label{Sec3}
In this section, we examine the behavior of the tripartite QM-EUR in the context of a Schwarzschild black hole. The incompatible observables measured on part $A$ of a three-qubit state $\rho_{ABC}$ are chosen as the Pauli matrices $X = \hat{\sigma}_{x}$ and $Z = \hat{\sigma}_{z}$.  Here, let us consider two different cases.
\\
\noindent\textbf{Case 1.} Alice, Bob, and Charlie are assumed to share a tripartite quantum state $\rho_{ABC}$ at the same initial point in flat Minkowski space-time. Particles $A$, $B$, and $C$ are sent to Alice, Bob, and Charlie, respectively. After receiving their particles, Alice remains stationary in an asymptotically flat region, while Bob and Charlie freely fall toward a Schwarzschild black hole and position themselves near the event horizon. Next, Alice performs either the $X$ or $Z$ measurement on her quantum system and communicates her measurement choice to Bob and Charlie. Bob's (or Charlie's) primary goal is to minimize his uncertainty about the $X$ ($Z$) measurement (see Fig. \ref{Schematic1}).
\\
\noindent\textbf{Case 2.} This scenario is similar to the previous one, but with the difference that Charlie and Bob remain in the asymptotically flat region, while Alice freely falls toward the black hole and positions herself near the event horizon (see Fig. \ref{Schematic2}).

Now, let's explore the above scenarios for two different initial states as mentioned before.

%%%%%%%%%%%%%%%%%%%%%%%%%%%%%%%%%
\begin{figure*}[!t]
\centering
\includegraphics[width=8cm]{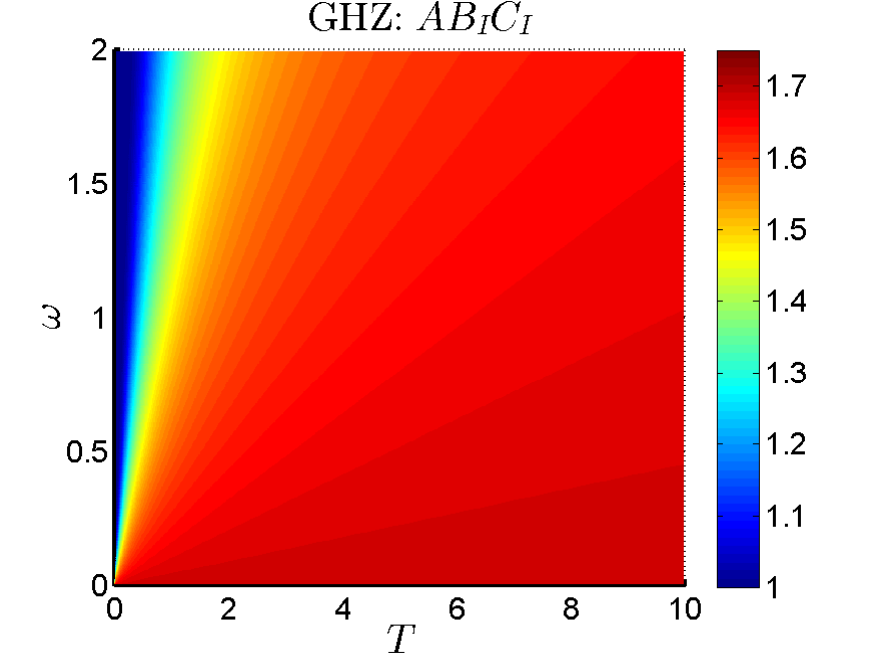}
\put(-220,150){{\bf (a)}}
\includegraphics[width=8cm]{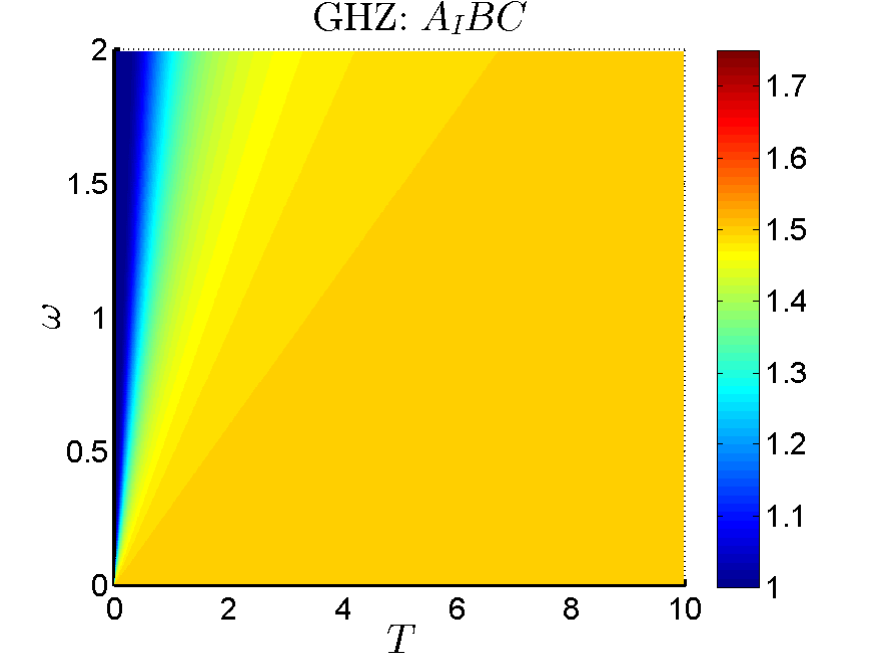}
\put(-220,150){{\bf(b)}}\\
\vspace{0.5cm}
\includegraphics[width=14.5cm]{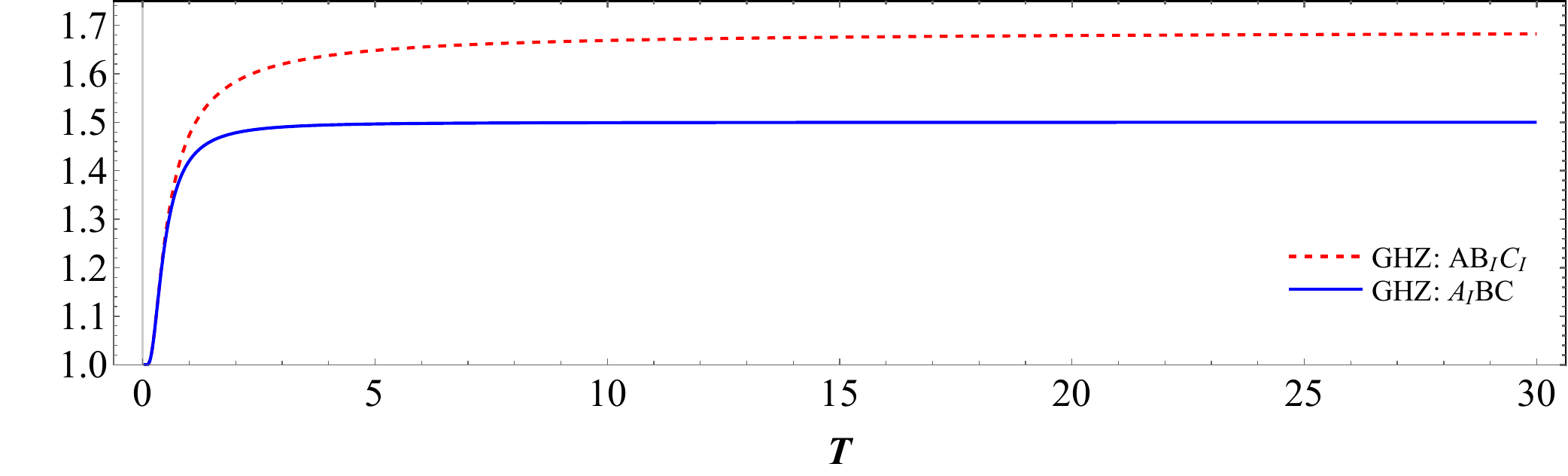}
\put(-415,115){{\bf(c)}}
\caption{Measurement uncertainties for both cases when the three qubits are initially prepared in the GHZ state. Plots {\bf(a)} and {\bf(b)}: $U$ versus $\omega$ and $T$. Plot {\bf(c)}: $U$ versus $T$ with $\omega=1$.}\label{fig1}
\end{figure*}
%%%%%%%%%%%%%%%%%%%%%%%%%%%%%%%%%

\subsection{GHZ state}
 The initial state of the system that has been shared between Alice, Bob, and  Charlie is assumed to be a GHZ state, given by
\begin{equation}\label{GHZ}
\vert \textmd{GHZ}\rangle_{A B C} = \dfrac{\vert 0\rangle_A \vert0\rangle_B \vert0\rangle_C + \vert 1\rangle_A \vert1\rangle_B \vert1\rangle_C}{\sqrt{2}}.
\end{equation}

For Case 1, by applying \eqref{bl1} and \eqref{bl2}, Eq. (\ref{GHZ}) can be re-expressed in terms of Minkowski modes for Alice and black hole modes for Bob and Charlie as follows:
\begin{align}
 |\textmd{GHZ}\rangle_{A B_{I} B_{II} C_{I} C_{II}}^{\textmd{Case 1}}=&\frac{1}{\sqrt{2}} \big[ \alpha |0\rangle_A |0\rangle_{B_{I}}|0\rangle_{B_{II}}|0\rangle_{C_{I}}|0\rangle_{C_{II}}\nonumber\\
&+\beta |0\rangle_A |1\rangle_{B_{I}}|1\rangle_{B_{II}}|1\rangle_{C_{I}}|1\rangle_{C_{II}} \nonumber\\
& +|1\rangle_A |1\rangle_{B_{I}}|0\rangle_{B_{II}}|1\rangle_{C_{I}}|0\rangle_{C_{II}}\nonumber\\
&+\alpha\beta |0\rangle_A |0\rangle_{B_{I}}|0\rangle_{B_{II}}|1\rangle_{C_{I}}|1\rangle_{C_{II}} \nonumber\\
& +\alpha\beta |0\rangle_A |1\rangle_{B_{I}}|1\rangle_{B_{II}}|0\rangle_{C_{I}}|0\rangle_{C_{II}}\big].
\end{align}
Since Region I is completely disconnected from Region II, Bob and Charlie cannot access the modes inside the event horizon. Thus, by tracing out the state of the inaccessible modes, the following density matrix can be obtained
\begin{equation}\label{GHZ2}
\rho^{\textmd{GHZ}}_{A B_I C_I}=\frac{1}{2}\left( \begin{array}{cccccccc}
\alpha^{4} & 0 & 0 & 0 & 0 & 0 & 0 & \alpha^{2} \\
0 & \alpha^{2} \beta^{2} & 0 & 0 & 0 & 0 & 0 & 0 \\
0 & 0 &  \alpha^{2} \beta^{2}  & 0 & 0 & 0 & 0 & 0 \\
0 & 0 & 0 &\beta^{4} & 0 & 0 & 0 & 0 \\
0 & 0 & 0 & 0 & 0 & 0 & 0 & 0 \\
0 & 0 & 0 & 0 & 0 & 0 & 0 & 0 \\
0 & 0 & 0 & 0 & 0 & 0 & 0 & 0 \\
\alpha^{2} & 0 & 0 & 0 & 0 & 0 & 0 & 1
\end{array}\right).
\end{equation}
Using Eqs. \eqref{tpu} and \eqref{GHZ2}, we obtain the analytical expression of  tripartite measurement uncertainty as follows
\begin{equation}\label{ughz1}
U^{\textmd{GHZ}}_{A B_I C_I}=\frac{-\eta  \ln \eta +\eta  \ln (2 (\eta +1))+\ln (2 \eta +2)+\theta  \ln 2}{\ln 4},
\end{equation}
with $\eta =\beta ^4+\alpha ^2 \beta ^2$ and $\theta =\alpha ^4+\alpha ^2 \beta ^2$.

In Case 2, where Charlie and Bob stay in the asymptotically flat region while Alice freely falls toward the black hole and is situated near the event horizon, Eq. (\ref{GHZ}) can be reformulated in terms of black hole modes for Alice and Minkowski modes for Bob and Charlie, namely
\begin{align}
|\textmd{GHZ}\rangle_{A_{I} A_{II} B C}^{\textmd{Case 2}}= &\frac{1}{\sqrt{2}}\big[\alpha |0\rangle_{A_{I}}|0\rangle_{A_{I I}}|0\rangle_{B} |0\rangle_{C} \nonumber\\
&+\beta |1\rangle_{A_{I}}|1\rangle_{A_{I I}}|0\rangle_{B} |0\rangle_{C}\nonumber\\
&+|1\rangle_{A_{I}}|0\rangle_{A_{I I}}|1\rangle_{B} |1\rangle_{C}\big].
\end{align}
Next, by tracing out the inaccessible mode $A_{I I}$, the following density matrix can be derived
\begin{equation}\label{GHZ3}
\rho^{\textmd{GHZ}}_{A_I B C}=\frac{1}{2}\left( \begin{array}{cccccccc}
\alpha^{2} & 0 & 0 & 0 & 0 & 0 & 0 & \alpha \\
0 & 0 & 0 & 0 & 0 & 0 & 0 & 0 \\
0 & 0 & 0 & 0 & 0 & 0 & 0 & 0 \\
0 & 0 & 0 & 0 & 0 & 0 & 0 & 0 \\
0 & 0 & 0 & 0 & \beta^{2} & 0 & 0 & 0 \\
0 & 0 & 0 & 0 & 0 & 0 & 0 & 0 \\
0 & 0 & 0 & 0 & 0 & 0 & 0 & 0 \\
\alpha & 0 & 0 & 0 & 0 & 0 & 0 & 1
\end{array}\right).
\end{equation}

Employing \eqref{tpu} and \eqref{GHZ3}, one can obtain the analytical form of  tripartite measurement uncertainty, given by
\begin{equation}\label{ughz2}
U^{\textmd{GHZ}}_{A_I B C}=\frac{-\alpha ^2\ln \frac{\alpha ^2}{2}-\beta ^2 \ln \frac{\beta ^2}{2}+\ln 2}{\ln 4}.
\end{equation}

In Fig. \ref{fig1}, the measurement uncertainties for the GHZ state are calculated based on Eqs. \eqref{ughz1} and \eqref{ughz2}. In Figs. \ref{fig1}(a) and \ref{fig1}(b), this uncertainty is plotted against the monochromatic frequency $\omega$ of the Dirac field and the Hawking temperature $T$ for two distinct scenarios: Case 1 and Case 2. Besides, Fig. \ref{fig1}(c) shows $U$ versus $T$ for $\omega = 1$.

According to all plots in Fig. \ref{fig1}, the measurement uncertainty $U$ increases monotonically with increasing Hawking temperature $T$. This indicates that the thermal effects due to Hawking radiation increase the uncertainty in measurements. These results are consistent with the findings of Ref. \cite{black9}, where it was demonstrated that higher Hawking temperatures lead to stronger Hawking radiation, which disturbs the quantum system more significantly. While the general trends are similar in Cases 1 and 2, the magnitude and rate of increase in $U$ with $T$ differ between the two cases, highlighting different levels of susceptibility to Hawking radiation depending on the specific arrangement of the particles.

If we consider the 2D plot in Fig. \ref{fig1}(c), one can analyze which of the two cases exhibits a higher $U$ for a fixed $\omega$. In Case 1, Bob and Charlie (quantum memories) are approaching the event horizon, while Alice (the measured particle) remains in the asymptotically flat region. Quantum memories near the event horizon are directly exposed to the intense gravitational effects and Hawking radiation. This exposure is expected to induce greater decoherence and entanglement degradation due to the strong interaction with the thermal radiation emanating from the black hole. Compared with Case 2, where Alice (the measured particle) is near the event horizon, while Bob and Charlie remain in the asymptotically flat region, Case 1 should feature a higher measurement uncertainty $U$ as a function of Hawking temperature $T$, which is demonstrated in Fig. \ref{fig1}.

In addition, we can discuss the impact of $\omega$ on the measurement uncertainty based on Fig. \ref{fig1}. In general, we notice that $U$ is reduced when $\omega$ increases. This effect is particularly pronounced at low Hawking temperatures. This suggests that higher frequency modes might mitigate some of the uncertainty introduced by the Hawking effect at lower temperatures.

At lower Hawking temperatures, the thermal radiation's effect is less significant. The higher frequency modes $\omega$ may help in reducing the uncertainty $U$ by interacting less destructively with the quantum memories.

%%%%%%%%%%%%%%%%%%%%%%%%%%%%%%%%%%%%
\begin{figure*}[!t]
\centering
\includegraphics[width=8cm]{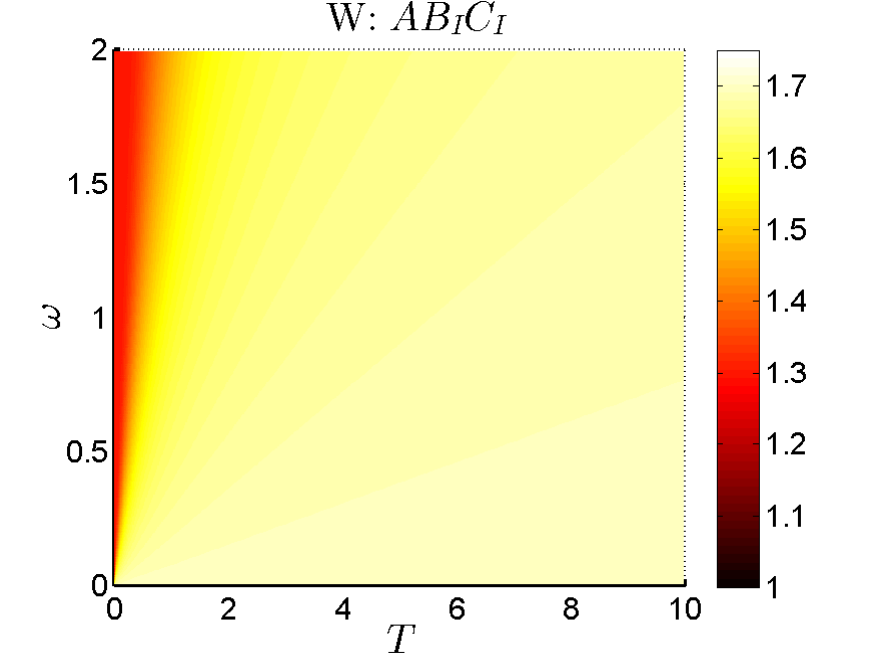}
\put(-220,150){{\bf (a)}}
\includegraphics[width=8cm]{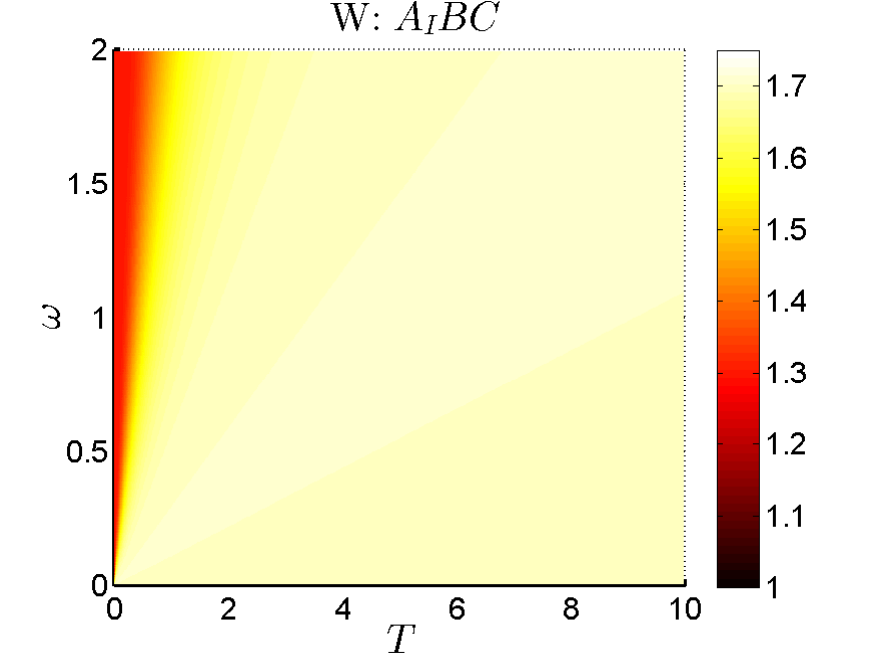}
\put(-220,150){{\bf (b)}}\\
\vspace{0.5cm}
\includegraphics[width=14.5cm]{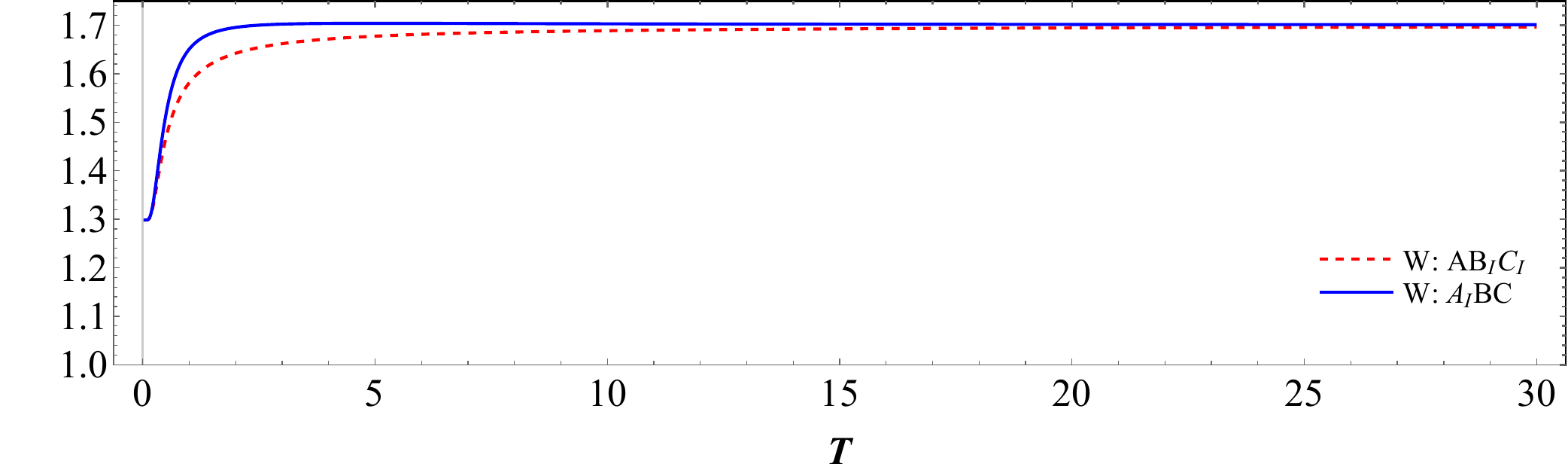}
\put(-415,115){{\bf(c)}}
\caption{Measurement uncertainties for both cases when the three qubits are initially prepared in the W state. Plots {\bf(a)} and {\bf(b)}: $U$ versus $\omega$ and $T$. Plot {\bf(c)}: $U$ versus $T$ with $\omega=1$.}\label{fig2}
\end{figure*}
%%%%%%%%%%%%%%%%%%%%%%%%%%%%%%%%%

\subsection{W state}
Let us assume that the W state shared by Alice, Bob, and Charlie is as follows
\begin{equation}\label{W}
\vert \textmd{W} \rangle_{A B C} = \dfrac{\vert 1\rangle_A \vert 0\rangle_B \vert 0\rangle_C + \vert 0\rangle_A \vert 1\rangle_B \vert 0\rangle_C + \vert 0\rangle_A \vert 0\rangle_B \vert 1\rangle_C}{\sqrt{3}}.
\end{equation}
The approach is similar to the previous section. Regarding Case 1, for the three qubits being prepared initially in the W state, Eq. (\ref{W}) can be rewritten as

 \begin{align}
\vert \textmd{W}\rangle_{A B_{I} B_{II} C_{I} C_{II}}^{\textmd{Case 1}}  =&\frac{1}{\sqrt{3}}\big[\alpha\vert 0\rangle_{A}\vert 0\rangle_{B_{I}} \vert 0\rangle_{B_{I I}} \vert 1\rangle_{C_{I}} \vert 0\rangle_{C_{I I}}\nonumber\\
&+\alpha\vert 0\rangle_{A} \vert 1\rangle_{B_{I}}  \vert 0\rangle_{B_{I I}} \vert 0\rangle_{C_{I}} \vert 0\rangle_{C_{I I}} \nonumber\\
& +\beta\vert 0\rangle_{A} \vert 1\rangle_{B_{I}} \vert 0\rangle_{B_{I I}} \vert 1\rangle_{C_{I}} \vert 1\rangle_{C_{I I}}\nonumber\\
&+\beta\vert 0\rangle_{A} \vert 1\rangle_{B_{I}} \vert 1\rangle_{B_{I I}} \vert 1\rangle_{C_{I}} \vert 0\rangle_{C_{I I}} \nonumber\\
& +\alpha^2\vert 1\rangle_{A} \vert 0\rangle_{B_{I}} \vert 0\rangle_{B_{I I}} \vert 0\rangle_{C_{I}} \vert 0\rangle_{C_{I I}}\nonumber\\
&+\alpha \beta\vert 1\rangle_{A} \vert 0\rangle_{B_{I}} \vert 0\rangle_{B_{I I}} \vert 1\rangle_{C_{I}} \vert 1\rangle_{C_{I I}} \nonumber\\
& +\alpha \beta\vert 1\rangle_{A} \vert 1\rangle_{B_{I}} \vert 1\rangle_{B_{I I}} \vert 0\rangle_{C_{I}} \vert 0\rangle_{C_{I I}}\nonumber\\
&+\beta^2\vert 1\rangle_{A} \vert 1\rangle_{B_{I}}  \vert 1\rangle_{B_{I I}} \vert 1\rangle_{C_{I}} \vert 1\rangle_{C_{I I}}\big].
\end{align}
Then, tracing over the inaccessible modes  $B_{II}$ and $C_{II}$, one comes to
\begin{equation}\label{W2}
\rho^{\textmd{W}}_{A B_I C_I}=\frac{1}{3}\left( \begin{array}{cccccccc}
0 & 0 & 0 & 0 & 0 & 0 & 0 & 0 \\
0 & \alpha^2 & \alpha^2 & 0 & \alpha^3 & 0 & 0 & 0 \\
0 & \alpha^2 & \alpha^2 & 0 & \alpha^3 & 0 & 0 & 0 \\
0 & 0 & 0 & 2 \beta^2 & 0 & \alpha \beta^2 & \alpha \beta^2 & 0 \\
0 & \alpha^3 & \alpha^3 & 0 & \alpha^4 & 0 & 0 & 0 \\
0 & 0 & 0 & \alpha \beta^2 & 0 & \alpha^2 \beta^2 & 0 & 0 \\
0 & 0 & 0 & \alpha \beta^2 & 0 & 0 & \alpha^2 \beta^2 & 0 \\
0 & 0 & 0 & 0 & 0 & 0 & 0 & \beta^4
\end{array}\right).\\
\end{equation}

Equipped with Eqs. \eqref{tpu} and \eqref{W2}, we arrive at
\begin{align}\label{uw1}
U^{\textmd{W}}_{A B_I C_I}=&-\frac{\alpha ^2+2 \beta ^2}{3} \log _2\frac{\alpha ^2+2 \beta ^2 }{3}-\frac{\Delta _-}{6} \log _2\frac{\Delta _- }{12}\nonumber\\
&+\frac{2 (\alpha ^2+\theta)}{3} \log _2\frac{\alpha ^2+\theta}{3}-\frac{\Delta _+}{6} \log _2\frac{\Delta _+ }{12}\nonumber\\
&-\frac{\alpha ^2}{3}  \log _2\frac{\alpha ^2}{3}-\frac{\theta}{3} \log _2\frac{\theta }{3}-\frac{\eta}{3} \log _2\frac{\eta }{3}\nonumber\\
&+\frac{\alpha ^2+2 \beta ^2+\eta}{3} \log _2\frac{\alpha ^2+2 \beta ^2+\eta}{3},
\end{align}
where $\Delta _{\pm}=3\pm\sqrt{16 \beta ^4-12 \beta ^2+5}$.

For Case 2, based on Eqs. (\ref{bl1}) and (\ref{bl2}), one can rewrite Eq. (\ref{W}) as follows:
 \begin{align}
|\textmd{W}\rangle_{A_{I} A_{II} B C}^{\textmd{Case 2}}  =&\frac{1}{\sqrt{3}}\big[\alpha \vert 0\rangle_{A_{I}} \vert 0\rangle_{A_{I I}} \vert 0\rangle_{B} \vert 1\rangle_{C}\nonumber\\
&+\beta\vert 1\rangle_{A_{I}} \vert 1\rangle_{A_{I I}} \vert 0\rangle_{B} \vert 1\rangle_{C} \nonumber\\
& +\alpha \vert 0\rangle_{A_{I}} \vert 0\rangle_{A_{I I}} \vert 1\rangle_{B} \vert 0\rangle_{C}\nonumber\\
&+\beta \vert 1\rangle_{A_{I}} \vert 1\rangle_{A_{I I}} \vert 1\rangle_{B} \vert 0\rangle_{C} \nonumber\\
& +\vert 1\rangle_{A_{I}} \vert 0\rangle_{A_{I I}} \vert 0\rangle_{B} \vert 0\rangle_{C} \big].
\end{align}
Tracing over the inaccessible region II,  one arrives at

\begin{equation}\label{W3}
\rho^{\textmd{W}}_{A_I B C}=\frac{1}{3}\left( \begin{array}{cccccccc}
0 & 0 & 0 & 0 & 0 & 0 & 0 & 0 \\
0 & \alpha^2 & \alpha^2 & 0 & \alpha & 0 & 0 & 0 \\
0 & \alpha^2 & \alpha^2 & 0 & \alpha & 0 & 0 & 0 \\
0 & 0 & 0 & 0 & 0 & 0 & 0 & 0 \\
0 & \alpha & \alpha & 0 & 1 & 0 & 0 & 0 \\
0 & 0 & 0 & 0& 0 & \beta^2 & \beta^2 & 0 \\
0 & 0 & 0 & 0 & 0 & \beta^2 & \beta^2 & 0 \\
0 & 0 & 0 & 0 & 0 & 0 & 0 & 0
\end{array}\right).
\end{equation}

Using now Eqs. \eqref{tpu} and \eqref{W3},  the following formula for tripartite uncertainty can be obtained
\begin{align}\label{uw2}
U^{\textmd{W}}_{A_I B C}=&-\frac{2 \alpha ^2}{3}  \log _2\frac{\alpha ^2}{3}-\frac{ \beta ^2}{3}  \log _2\frac{\beta ^2}{3}\nonumber\\
&-\frac{ \Gamma _-}{6}  \log _2\frac{\Gamma _-}{12}-\frac{ \Gamma _+}{6}  \log _2\frac{\Gamma _+}{12}\nonumber\\
&+\frac{ 2}{3}  \log _2\frac{1}{3}+\frac{ 4}{3}  \log _2\frac{2}{3}\nonumber\\
&-\frac{ 1+\beta ^2}{3}  \log _2\frac{1+\beta ^2}{3},
\end{align}
where $\Gamma _{\pm }=3\pm \sqrt{4 \alpha ^2+1}$.

%%%%%%%%%%%%%%%%%%%%%%%%%%%%%%%%%%%%%%
\begin{figure}[t]
\centering
\includegraphics[width=8cm]{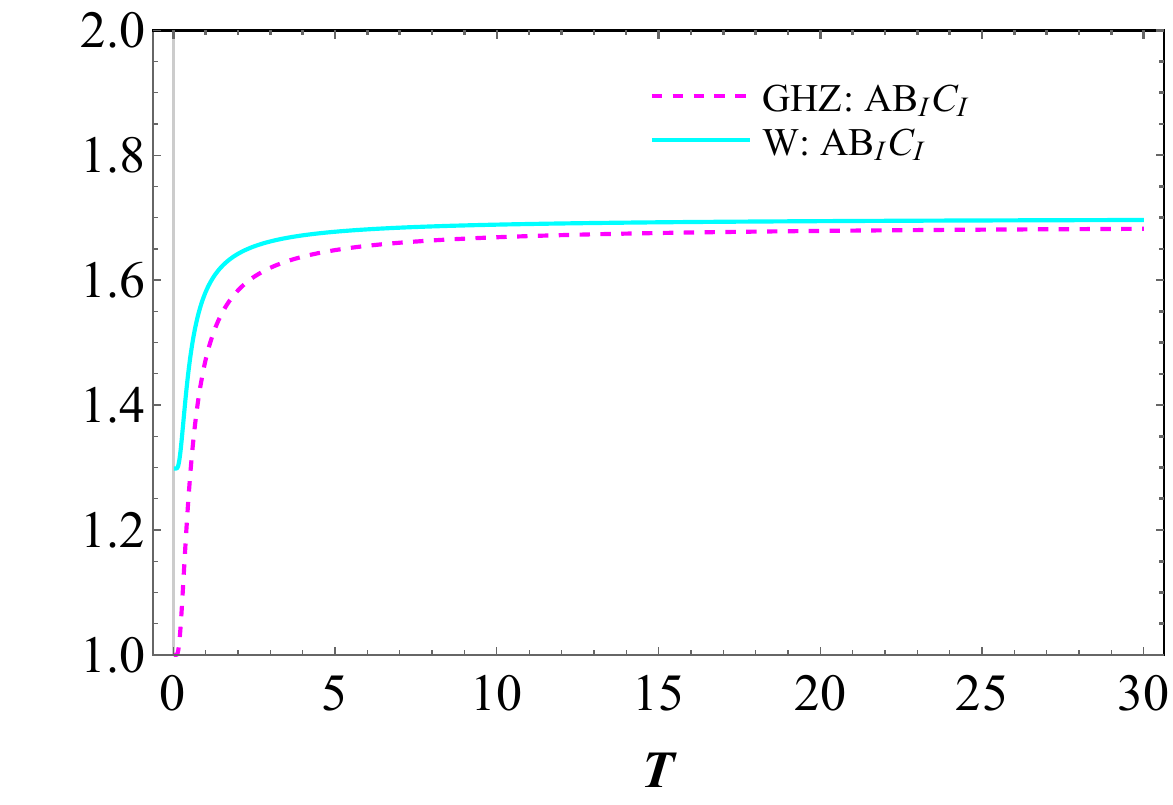}
\caption{Measurement uncertainty for W and GHZ states versus the Hawking temperature $T$, with Alice remaining stationary in an asymptotically flat region while Bob and Charlie freely fall toward a Schwarzschild black hole, where $\omega=1$.}\label{fig3}
\end{figure}
%%%%%%%%%%%%%%%%%%%%%%%%%%%%%%%%%%%%%%%

%%%%%%%%%%%%%%%%%%%%%%%%%%%%%%%%%%%%%%
\begin{figure}[t]
\centering
\includegraphics[width=8cm]{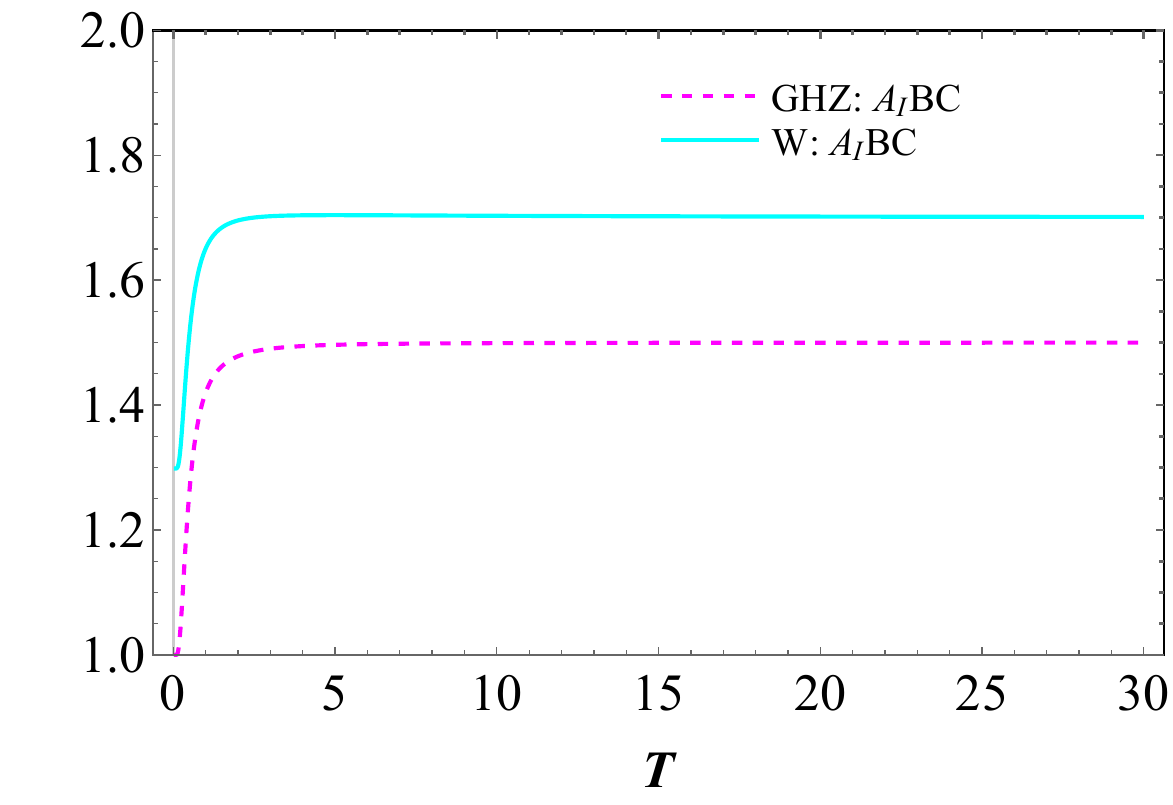}
\caption{Measurement uncertainty for W and GHZ states versus the Hawking temperature $T$, with Charlie and Bob staying in the asymptotically flat region while Alice falls freely toward the black hole, where $\omega=1$.}\label{fig4}
\end{figure}
%%%%%%%%%%%%%%%%%%%%%%%%%%%%%%%%%%%%%%

The measurement uncertainties for the W state [see \eqref{uw1} and \eqref{uw2}] are plotted as functions of the monochromatic frequency $\omega$ of the Dirac field and the Hawking temperature $T$ in Figs.
\ref{fig2}(a) and \ref{fig2}(b). Similar to the GHZ state, for both cases, the measurement uncertainty $U$ decreases with increasing $\omega$ at low temperatures. The decline of $U$ for higher values of $\omega$ at low temperatures again suggests that higher frequency modes are less disruptive to the W state’s entanglement.

Moreover, measurement uncertainty $U$ increases monotonically with increasing the Hawking temperature $T$. As $T$ increases, the Hawking radiation’s thermal effects dominate, leading to increased measurement uncertainty. When we compare the two cases, we see that the general trends are similar, but the magnitude and rate of change in $U$ with respect to $T$ and $\omega$ differ, reflecting different susceptibilities to Hawking radiation depending on the position of the quantum memories and the measured particle.

In Fig. \ref{fig2}(c), we also present $U$ as a function of $T$ with a fixed $\omega$. Our observation indicates that the measurement uncertainty behaves differently for the W state compared to the GHZ state at lower temperatures. Specifically, for the W state, Case 1 exhibits less uncertainty at lower temperatures, but as the temperature increases, the uncertainties for both cases tend to converge to the same value.

At lower Hawking temperatures, the effect of thermal radiation is minimal. The robustness of the W state to particle loss means that the entanglement and coherence of the system are better preserved, resulting in lower measurement uncertainty for Case 1.

\subsection{Comparison between GHZ and W states}
To compare the GHZ state with the W state, in Figs. \ref{fig3} and \ref{fig4} dependence of the measurement uncertainty on the Hawking temperature $T$ is plotted, where Fig. \ref{fig3} presents Case 1
and Fig. \ref{fig4} illustrates Case 2.

For Case 1 presented in Fig. \ref{fig3}, the measurement uncertainty starts at a lower value for the GHZ state compared to the W state. This indicates that the GHZ state is initially less affected by the Hawking radiation at low temperatures, maintaining better coherence. As the Hawking temperature $T$ continues to increase, the measurement uncertainties for both the W and GHZ states converge to similar values. This convergence indicates that at high temperatures, the overwhelming thermal effects of the Hawking radiation uniformly disrupt both types of quantum states, making the initial differences in their entanglement properties less significant.

In Fig. \ref{fig4}, the initial measurement uncertainty for the W state is higher, suggesting that the W state is more sensitive to the initial presence of Hawking radiation, even at low temperatures. As the Hawking temperature $T$ increases, the uncertainties for the GHZ and W states increase monotonically. Despite this increase, the GHZ state maintains a lower uncertainty compared to the W state across all temperatures. At high Hawking temperatures, the measurement uncertainty for the GHZ state approaches an asymptotic value that is lower than that for the W state. This indicates that even at high temperatures, the GHZ state retains better coherence and lower uncertainty compared to the W state.

The difference between the GHZ and W states can be further understood by considering the geometry of their entanglement. In the GHZ state, all the entanglement is shared globally among all three qubits, meaning the entanglement is strictly three-way \cite{Miyake2003,Yu2008}. This monogamous nature of the GHZ state ensures that the entanglement is concentrated, creating a strong, unified quantum correlation that is resistant to external disturbances like Hawking radiation. In contrast, the W state's entanglement is more distributed, allowing two-qubit subsystems to retain some level of entanglement even if one qubit is lost. However, this lack of strong three-way entanglement means the W state does not have the same collective defense against decoherence. As a result, the more dispersed entanglement in the W state leaves it more susceptible to gradual disruptions, such as those caused by low levels of Hawking radiation, leading to higher measurement uncertainty. The GHZ state's three-way entanglement geometry, being more tightly bound, helps it maintain coherence more effectively across a range of temperatures.

The research presented in this paper aligns closely with the findings of S.-M. Wu et al. \cite{black10}, particularly in demonstrating the superior robustness of the GHZ state against the Hawking effect compared to the W state. Both studies show that as the Hawking temperature increases, the GHZ state's entanglement properties exhibit greater resilience. In our work, this is reflected in the lower initial measurement uncertainty and the less steep increase in uncertainty for the GHZ state, even as temperature rises. This consistency strengthens the argument that the GHZ state has inherent advantages in maintaining quantum coherence in extreme conditions such as near a black hole's event horizon.

However, the approach used in the present contribution is original and novel in several key aspects. While Wu et al. \cite{black10} focused on the behavior of genuine tripartite entanglement (GTE) and tangle measures, our research specifically analyzes the measurement uncertainty $U$, which offers a different perspective on quantum state robustness. By investigating how $U$ changes under various conditions---such as different positions of quantum memories and measured particles relative to the event horizon---we provide a more comprehensive view of quantum state behavior in Schwarzschild spacetime. This novel approach not only validates previous findings about the GHZ state's resilience but also extends our understanding of the impact of Hawking radiation on quantum measurements, thereby contributing valuable insights to the field of relativistic quantum information processing.

%\vspace{2cm}
%%%%%%%%%%%%%%%%%%%%%%%%%%%%%%%%%%%%%%%%%%
\section{Conclusion}\label{conclusion}
Several studies have investigated quantum correlations in a tripartite system within the context of a Schwarzschild black hole, revealing that their dynamical behaviors are significantly influenced by the Hawking temperature $T$ \cite{black9,black10,new1,new2,new3}. These studies demonstrated that the Hawking effect diminishes quantum correlations. Since measurement uncertainty in a QM-EUR is closely linked to the system's quantum correlations, it is anticipated that it too may be impacted by the Hawking temperature $T$. In this work, the effect of Hawking radiation on the tripartite QM-EUR in Schwarzschild space-time was studied for GHZ and W states. It has been shown that the behaviors of uncertainty depend on the Hawking effect. Specifically, it has been found that the uncertainty increases monotonically with increasing Hawking temperature. As the Hawking temperature $T$ increases, the intensity of Hawking radiation also increases, leading to greater decoherence and increased measurement uncertainty. However, the higher monochromatic frequency $\omega$ of the Dirac field might mitigate some of the uncertainties introduced by the Hawking radiation.

As for the comparison between GHZ and W states, the GHZ state starts with a lower measurement uncertainty at low Hawking temperatures compared to the W state. This indicates that the GHZ state is initially more resilient to the effects of Hawking radiation. Additionally, in the scenario where Charlie and Bob remain in the asymptotically flat region and Alice falls toward the black hole, the uncertainties for the GHZ and W states do not converge at high temperatures. The GHZ state consistently maintains a lower measurement uncertainty than the W state, highlighting its superior robustness against Hawking radiation.

These findings contribute to a deeper understanding of quantum mechanics in black hole environments and could have implications for quantum information processing and communication in extreme conditions.

\vspace{1cm}
\noindent \textbf{Acknowledgments:}
H.D. acknowledges the support of project VEGA 2/0183/21 (DESCOM). \\
\noindent \textbf{Data availability:} No datasets were generated or analyzed during the study.\\
\noindent \textbf{Code availability:} This manuscript has no associated code/software.\\
\noindent \textbf{Competing interests:} The authors declare no competing interests.

\newpage

%=============================================================%
%=============================================================%
%=======================  References =========================%
%=============================================================%
%=============================================================%

%%%%%%%%%%%%%%%%%%%%%%%%%%%%%%%%%%%%%%%%%%%%%%%%%%%%%%%%%%%%%%%%%%%%%
%%%%%%%%%%%%%%%%%%%%%%%%%%%%%%%%%%%%%%%%%%

%%%%%%%%%%%%%%%%%%%%%%%%%%%%%%%%%%%%%%%%%%%%%%%%%%%%%%%%%%%%%%%%%%%%%
%%%%%%%%%%%%%%%%%%%%%%%%%%%%%%%%%%%%%%%%%%%%%%%%%%%%%%%%%%%%%%%%%%%%

\end{document}